\documentclass[article,prb,twocolumn,floatfix,showpacs]{revtex4-1}
\usepackage{epsfig,subfigure,amssymb,amscd,amsmath,graphicx,amsfonts}
\usepackage{times,longtable}

\newcommand{\non}{\nonumber}

\newcommand{\bea}{\begin{eqnarray}}
\newcommand{\eea}{\end{eqnarray}}
\newcommand{\be}{\begin{equation}}
\newcommand{\ee}{\end{equation}}
\newcommand{\ba}{\begin{align}}
\newcommand{\ea}{\end{align}}

\newcommand{\ket}[1]{     |    \,    #1    \rangle}
\newcommand{\bra}[1]{  \langle #1  \,  |}

\renewcommand{\vr}{{\boldsymbol{r}}}

\usepackage[usenames]{color}

\begin{document}

\title{Endstates in multichannel spinless $p$-wave superconducting wires}

\author{M.-T. Rieder, G. Kells, M. Duckheim, D. Meidan, P. W. Brouwer}

\affiliation{Dahlem Center for Complex Quantum Systems and Fachbereich Physik, Freie Universit\"{a}t Berlin, Arnimallee 14, 14195 Berlin, Germany}

\begin{abstract}
Multimode spinless $p$-wave superconducting wires with a width $W$ much smaller than the superconducting coherence length $\xi$ are known to have multiple low-energy subgap states localized near the wire's ends. Here we compare the typical energies of such endstates for various terminations of the wire: A superconducting wire coupled to a normal-metal stub, a weakly disordered superconductor wire, and a wire with smooth confinement. Depending on the termination, we find that the energies of the subgap states can be higher or lower than for the case of a rectangular wire with hard-wall boundaries.
\end{abstract}


\date{\today} \maketitle

\section{Introduction}

In the current search for Majorana fermions in nano-wire geometries \cite{Beenakker2011,Alicea2012} an important theoretical challenge is to understand the multiplicity of possible fermionic bound states that can form at the ends of the wire and how a possible
Majorana bound state can be identified among them. This is particularly relevant for multichannel geometries, in which fermionic states localized near the ends of the wire are expected to occur at energies much smaller than the excitation gap for bulk excitations if the wire width is much smaller than the superconducting coherence length. In this article we explore the dependence of these sub-gap end-states on the details of the termination of the wire and on impurity scattering.

The interest in isolating Majorana fermions arises because their non-local properties and non-abelian braiding statistics render them potentially useful for fault tolerant quantum computation.\cite{kitaev2003,freedman1998,Read2000,Ivanov2001,Kitaev2006,Nayak2008,Wilczek2009} Majorana fermions occur --- at least theoretically --- at the ends of one-dimensional spinless $p$-wave superconductors.\cite{Kitaev2001} Recent proposals suggest ways of engineering solid-state systems that effectively behave as spinless $p$-wave superconductors by combining an s-wave superconductor and a topological insulator,\cite{Fu2008,Cook2011} a semiconductor \cite{Sau2010,Alicea2010,Oreg2010,Lutchyn2010} or ferromagnet.\cite{Duckheim2011,Chung2011,Choy2011,Martin2012,Kjaergaard2012} Building on the proposals of Refs.\ \onlinecite{Oreg2010,Lutchyn2010}, two experimental groups have reported an enhanced tunneling density of states at the ends of InAs and InSb wires in proximity to a superconductor, consistent with the existence of Majorana bound states at the ends of these wires,\cite{Mourik2012,Das2012} whereas a number of other groups claim the observation of Majorana bound states using different methods.\cite{Williams2012,Rokhinson2012,Deng2012}

Whereas the original proposals for Majorana fermions in wire geometries focused on one-dimensional systems, 
it is by now well established that the topological superconducting phase with Majorana end states may persist 
in a quasi-one-dimensional multichannel 
setting.\cite{Wimmer2010,Potter2010,Lutchyn2011,Potter2011,Stanescu2011,Zhou2011,Kells2012,Potter2012,Lim2012,Gibertini2012,Tewari2012} 
A difference between the quasi-one-dimensional and one-dimensional settings is, however, that a possible 
zero-energy Majorana state localized at the wire's end may co-exist with other fermionic sub-gap states,
analogous to those found in vortex cores of bulk superconductors.\cite{Caroli1964} For the case of an 
$N$-channel spinless $p+ip$ superconductor with a rectangular geometry and with width $W$ much smaller 
than the superconducting coherence length $\xi$, three of us recently showed that the number of such 
fermionic subgap states is $\sim N/2$, and that their typical energy is $\varepsilon_{\rm typ} \sim 
\Delta (W/\xi)^2$, $\Delta$ being the superconducting gap size.\cite{Kells2012} The lowest-lying and 
highest-lying fermionic subgap states have energies 
$\varepsilon_{\rm min} \sim \varepsilon_{\rm typ}/N \ln N$ and $\varepsilon_{\rm max} \sim N 
\varepsilon_{\rm typ}$, respectively. The fermionic subgap states also exist in a non-topological 
phase without zero-energy Majorana end-state, thus posing a potential obstacle for the identification 
of the topological phase through the observation of an enhanced density of states near zero energy.

In a recent article, Potter and Lee\cite{Potter2012} observe that the dependence of the energy of the 
lowest-lying fermionic subgap state on system parameters changes qualitatively if the rectangular 
geometry of Ref.\ \onlinecite{Kells2012} is replaced by a geometry with rounded ends. They point out 
that the calculation of the energy of the fermionic subgap state for the rectangular geometry is plagued 
by a subtle cancellation, which does not appear for a generic wire ending. In particular it was found in Ref.~\onlinecite{Potter2012} 
that the lowest-lying fermionic subgap state has an energy significantly above the prediction of 
Ref.\ \onlinecite{Kells2012} for a wire with width $W \sim \xi$ and rounded ends.

Motivated by these observations we present here a detailed investigation of 
the effect that the wire termination has on the energies of the fermionic subgap states for the multichannel 
spinless $p + i p$ superconductor. Remarkably, we find that, depending on the details of the wire ending, 
the energies of the fermionic subgap states can be significantly above, as well as below the 
rectangular-wire case of Ref.\ \onlinecite{Kells2012}. We find an increase of the energies of the 
subgap states if an arbitrarily-shaped normal layer is attached to the wire's end, the magnitude 
of the increase being
consistent with the estimate of Ref.\ \onlinecite{Potter2012} for a wire with rounded ends.
On the other hand, the presence of impurities --- weak enough to preserve the topological phase \cite{Motrunich2001,Brouwer2011b} --- on
average {\em reduces} the energies of the fermionic end states below the estimate of Ref.\ 
\onlinecite{Kells2012}, while a smooth confinement (with a slowly increasing potential energy
providing the confinement along the wire's axis) leads to even smaller energies of the fermionic 
subgap states.

Our results are derived for the two-dimensional spinless $p+ip$ superconducting strip of width $W$. The model
of a spinless $p + i p$ superconductor is an effective low-energy description for the various proposals 
to realize one-dimensional or quasi-one-dimensional topological superconductors, provided the number 
$N$ of propagating channels at the Fermi level is chosen equal to the number of spinless (i.e., 
helical or spin-polarized) channels in the case of the semiconductor or ferromagnet proposals. (The 
edges of a topological insulator always have $N=1$, so that a multichannel $p + ip$ model is not
relevant in that case.) A mapping between the spinless $p$-wave model and the 
semiconductor-wire proposals is given in the appendix.

The remainder of this article is organized as follows: In Sec.\ \ref{sec:2} we briefly review the 
symmetries of the model (\ref{eq:HBdG}) and the reason for the appearance of multiple low-lying 
states if the wire width $W$ is much smaller than the superconducting coherence length $\xi$. 
In Sec.\ \ref{sec:3} we describe a scattering theory of fermionic subgap states with arbitrary wire endings. 
Section \ref{sec:4} discusses the $p+ip$ model with weak disorder, while the effect of a smooth potential 
at the wire's end is discussed in Sec.\ \ref{sec:5}. We conclude in Sec.\ \ref{sec:6}.
  
\section{$p+ip$ model} \label{sec:2}

Our calculations are performed for a two-dimensional spinless $p+ip$ superconductor, which is described by the two-component Bogoliubov-de Gennes Hamiltonian, which we write as
\be
  H = H_0 + H_y + H_V,
\ee
with
\begin{eqnarray}
  H_0 &=& \left( \frac{p^2}{2 m} - \mu \right) \tau_z +
  \Delta' p_x \tau_x, \nonumber \\
  H_y &=& - \Delta' p_y \tau_y, \nonumber \\
  H_V &=& V(\vr) \tau_z.
  \label{eq:HBdG}
\end{eqnarray}
Here $\tau_x$, $\tau_y$, and $\tau_z$ are Pauli matrices in particle-hole space, $\Delta'$ specifies the $p$-wave superconducting order parameter, $\mu = \hbar^2 k_{\rm F}^2/2m$ and $m$ are the 
chemical potential and electron 
mass, and $V(\vr)$ a potential that describes the confinement at the ends of the wire as well as the 
scattering off impurities. The two-dimensional coordinate $\vr=(x,y)$, where $0 < y < W$, with hard-wall 
boundary conditions at $y=0$ and $y=W$. The superconducting order parameter derives from 
proximity coupling to a bulk superconductor, so that no self-consistency condition for $\Delta'$ needs to be employed. 

Hypothetical end states are localized within a distance of the order of the superconducting 
coherence length $\xi = \hbar (\Delta' m)^{-1}$ from the wire's ends. For thin wires with $W \ll \xi$ it is a good starting point to analyze 
the Hamiltonian $H = H_0 + H_V$ without the term $H_y$. The Hamiltonian $H_0$ has a
chiral symmetry,\cite{Tewari2011} $\tau_y H_0 \tau_y = -H_0$, and there exist 
\be
  N = \mbox{int}\, [(W/\pi)\sqrt{k_{\rm F}^{2} - \xi^{-2}}]
\ee
Majorana bound states at each end of the wire. \cite{Kells2012,Potter2012,Gibertini2012,Tewari2012}
The stepwise increase of the number of Majorana end states for wire widths $W$ such 
that $(W/\pi)\sqrt{k_{\rm F}^{2} - \xi^{-2}}$ is an integer is accompanied by a closing of the bulk excitation gap of $H_0$. Inclusion of the potential term $H_V$ does not lift the degeneracy of the Majorana end states, since $H_V$ preserves the chiral symmetry, although it may change the boundaries of the phases with different $N$ if $H_V$ is nonzero in the bulk of the wire. In contrast, the term $H_y$ breaks the chiral symmetry and couples the $N$ Majorana bound states,
giving rise to (generically) $\mbox{int}\, (N/2)$ fermionic states at each end and a single Majorana end
state if $N$ is odd. If $W \ll \xi$ the splitting of the end states is small in
comparison to the bulk energy gap $\Delta = \Delta' \hbar k_F $, and the resulting fermionic states cluster near zero energy.\cite{Kells2012,Potter2012}

\begin{figure}
\centering
\includegraphics[width=.5\textwidth]{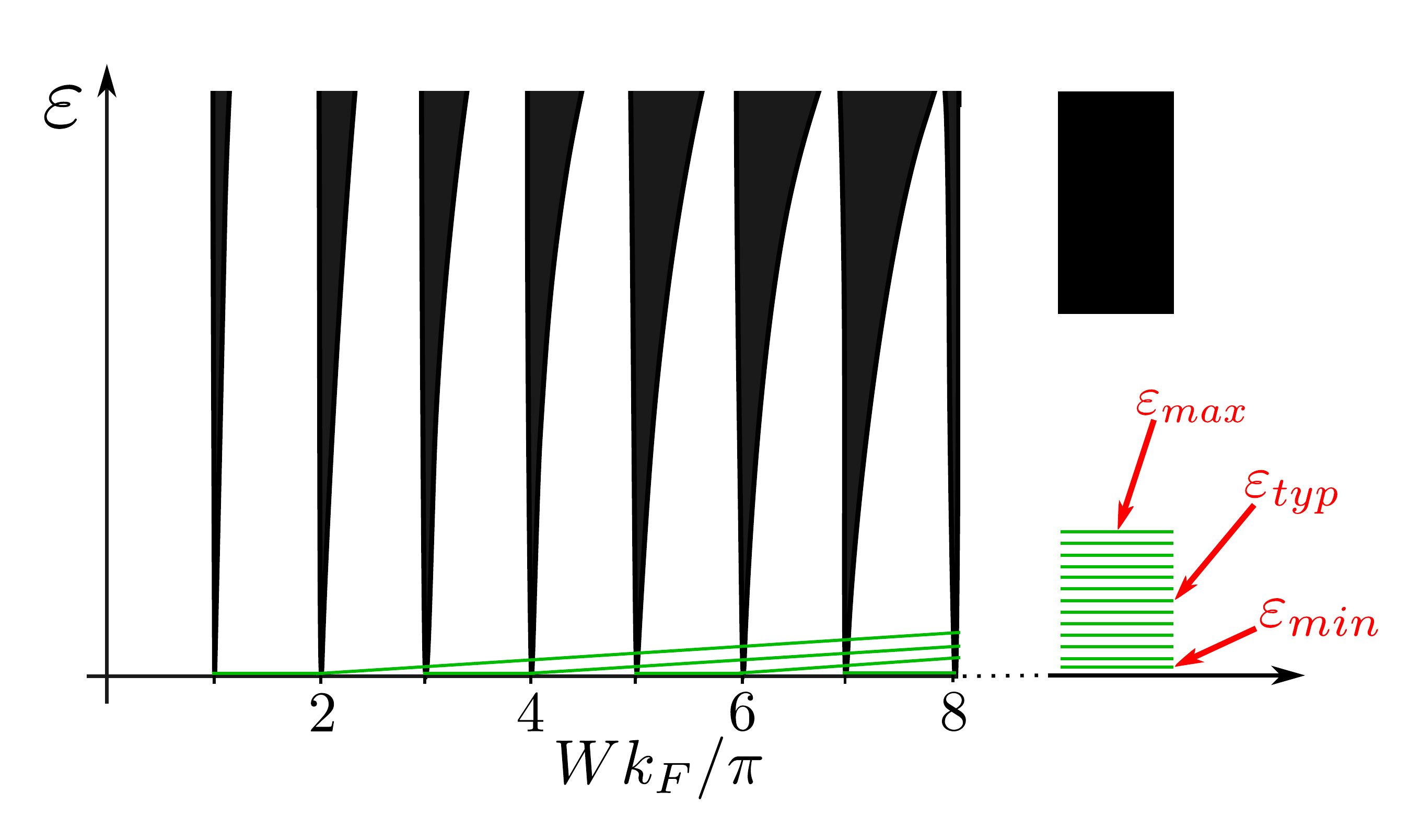}
\caption[]{(Color online) Schematic picture of the spectrum of low-energy excitations of the a $p+ip$ wire as a function of its width $W$. The gap for bulk excitations closes at those values of $W$ for which $(W/\pi)\sqrt{k_{\rm F}^{2} - \xi^{-2}}$ is an integer. When the bulk gap is finite, there are low-energy subgap states localized near the ends of the wire. In the text, we use $\varepsilon_{\rm min}$ to denote the energy of the lowest-lying fermionic subgap state, $\varepsilon_{\rm typ}$ for the typical energy of a subgap state, and $\varepsilon_{\rm max}$ for the energy of the highest-lying fermionic subgap state.
}
\label{fig:1}
\end{figure}

A schematic picture of the end-state spectrum as a function of $W$ is shown in Fig.\ \ref{fig:1}. The
end states are characterized by the energy $\varepsilon_{\rm min}$ of the lowest-lying fermionic end
state, the typical end-state energy $\varepsilon_{\rm typ}$, and the energy $\varepsilon_{\rm max}$ of
the highest-lying end state. For small $N$ these three energy scales are comparable, but for large $N$
they may differ considerably. The energy $\varepsilon_{\rm min}$ serves as the ``energy gap'' protecting
the topological state and sets the required energy resolution if the presence or absence of a Majorana
end state is detected through a tunneling density of states measurement.

The specific case of a rectangular wire geometry, with hard-wall boundary conditions at each end of the 
wire and without disorder, was investigated in Ref.\ \onlinecite{Kells2012}. We now investigate two other possible terminations, as well as the effect of disorder on the energies of subgap endstates in multichannel spinless $p$-wave superconducting wires.

\section{Normal-metal stub}
\label{sec:3}

\begin{figure}
\centering
\includegraphics[width=.3\textwidth]{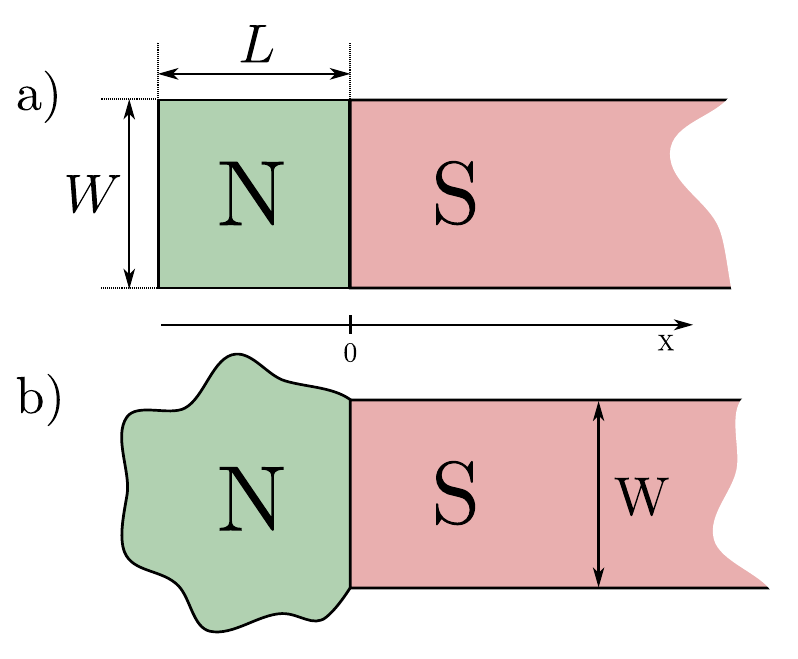}
\caption[]{(Color online) Schematic drawing of a spinless $p$-wave superconducting wire (S) coupled to a normal-metal (N) stub at one end. The top panel shows a rectangular stub, the bottom panel shows a chaotic cavity attached to the superconducting wire. }
\label{fig:3}
\end{figure}

In this section, we consider a quasi-one-dimensional spinless $p + ip$ superconductor without disorder and coupled to a normal-metal stub at its end. We choose coordinates, such that the spinless superconductor occupies the space $x > 0$, $0 < y < W$, see Fig.\ \ref{fig:3}. Such a wire ending is relevant, {\em e.g.}, for the experimental geometry of Ref.\ \onlinecite{Mourik2012}, in which a topological phase is induced in a semiconductor nanowire by laterally coupling it to a superconductor, while a part of the wire sticks out from under the superconductor and is pinched off by a gate at a finite distance. 

We take the Hamiltonian of the normal stub to be real and symmetric, in order to preserve the chiral symmetry of the Hamiltonian $H_0$. Following Ref.\ \onlinecite{Kells2012} we first solve for the wavefunctions $\psi^{(j)}$ of the $N$ Majorana modes for the Hamiltonian $H_0$ and then treat $H_y$ in perturbation theory. The potential term $H_V$ is set to zero throughout this calculation.

The Majorana states have support in the normal stub as well as in a segment of the superconducting wire of length $\sim \xi$. In the superconducting region $x > 0$ the wavefunctions $\psi$ of the Majorana states can be written as
\begin{eqnarray} 
  \label{eq:expan_major}
  \psi(\vr) &=& \sum_{n} a_{n-} \phi_{n-}(\vr) + a_{n+} \phi_{n+}(\vr), 
\end{eqnarray}
where the basis states $\phi_{n\pm}$, $n=1,2,\ldots,N$, read
\begin{eqnarray}
  \phi_{n\pm}(\vr) &=& \left( \begin{array}{c} e^{i\pi/4}\\ e^{-i\pi/4} \end{array} \right) 
  \sqrt{\frac{2 m}{W \hbar k_n}}
  e^{\pm i k_n x - x/\xi} \sin \left(\frac{n\pi y}{W}\right), \nonumber \\
  \label{eq:spinor}
\end{eqnarray}
with
\be
  k_{n} = \sqrt{k_{\rm F}^2 - (n \pi/W)^2}.
  \label{eq:kn}
\ee
The basis states $\phi_{n\pm}$ have been normalized to unit flux. The above expressions for the basis states and their normalization are valid up to corrections of order $(W/\xi)^2$, which we neglect throughout this calculation.

The coupling to the normal-metal stub imposes boundary conditions on the coefficients $a_{n\pm}$, which we express in terms of the scattering matrix $S_{nn'}$ of the normal stub,
\be
  a_{n+} = \sum_{n'} S_{nn'} a_{n'-},\ \
  a_{n-} = \sum_{n'} S_{nn'}^* a_{n'+},
  \label{eq:aeq}
\ee
Because the Hamiltonian of the normal stub is real and symmetric, the scattering matrix $S_{nn'}$ is unitary and symmetric, $S_{nn'} = S_{n'n}$, which ensures that the $2N$ equations (\ref{eq:aeq}) have $N$ independent solutions, corresponding to the $N$ Majorana end states. 

For finding an explicit representation of the $N$ Majorana states $\psi^{(j)}$, $j=1,2,\ldots,N$ we use the fact that the scattering matrix $S$ and the Wigner-Smith
time-delay matrix\cite{Wigner1955,Smith1960} $Q = i \hbar S^{\dagger} \partial S/\partial \mu$ of the 
normal stub can be simultaneously decomposed as
\be
  S = U^{\rm T} U,\ \ Q = U^{\dagger} \mbox{diag}\, (\tau_1,\ldots,\tau_N) U,
\ee
where $U$ is an $N \times N$ unitary matrix and the $\tau_i > 0 $, $i=1,2,\ldots,N$, are the
so-called ``proper time delays''. With this decomposition,
a solution to the boundary conditions (\ref{eq:aeq}) is given by
\be
  a_{n+}^{(j)} = U_{nj},\ \
  a_{n-}^{(j)} = U_{nj}^*,\ \
  j=1,2,\ldots,N.
\ee
The $N$ states that are defined through these coefficients,
\begin{eqnarray} 
  \label{eq:expan_major2}
  \tilde \psi^{(j)}(\vr) &=& \sum_{n} a_{n-}^{(j)} \phi_{n-}(\vr) + a_{n+}^{(j)} \phi_{n+}(\vr), 
\end{eqnarray}
are Majorana modes (they satisfy $\tilde \psi^{(j)*} = \tau_x \tilde \psi^{(j)}$), but they are not necessarily orthonormal. In order to construct an orthonormal set, we first calculate the scalar product $M_{jl}$ of the modes $\tilde \psi^{(j)}$,
\begin{eqnarray}
  M_{jl} &=&
  \int_0^{\infty} dx \int_0^W dy\, \tilde \psi^{\ast(j)}(\vr) \tilde \psi^{(l)}(\vr) 
  \nonumber \\ && \mbox{}
  + \int_{\text{stub}} d\textbf{r} \tilde \psi^{\ast(j)}(\vr) \tilde \psi^{(l)}(\vr)
  \\ &=&
  \sum\limits_{n} \frac{2 m \xi}{\hbar k_n} 
  \mbox{Re}\, \left( U_{nj} U^{*}_{nl} + \frac{ U_{nj} U^{*}_{nl}}{1 - i k_{n} \xi} \right)
  + 2 \tau_{j} \delta_{jl}, \nonumber
\end{eqnarray}
Here we used the relation between the Wigner-Smith time delay matrix and the normalization of scattering states in order to perform the integration over the sub, see Ref.\ \onlinecite{Smith1960}.
The overlap matrix $M$ is real, positive definite, and symmetric. It is manifestly diagonal if the scattering matrix $S$ and the time-delay matrix $Q$ are diagonal or in the ``large-stub limit'', which is defined as the limit in which the mean inverse dwell time $\hbar/\bar \tau$ is much smaller than the superconducting gap. In both cases, one obtains an orthonormal basis for the $N$ Majorana modes by setting $\psi^{(j)} = \tilde \psi^{(j)}/\sqrt{M_{jj}}$. In the general case, $M$ is not diagonal, however, and one has to construct an orthonormal system with the help of the orthogonal transformation $O$ that diagonalizes $M$, {\em i.e.}, $ O^{T}MO = \lambda^2$, where $\lambda = \mbox{diag}\, (\lambda_1,\lambda_2,\ldots,\lambda_N)$ is a diagonal matrix with positive elements. The corresponding orthonormal basis set one thus obtains reads
\begin{align} \label{eq:trafo_orthonorm}
  \psi^{(j)}(\vr) = 
  \sum_{n=1}^{N} \tilde \psi^{(n)}(\vr) O_{nj} \lambda_n^{-1}.
\end{align}

Inclusion of $H_y$, which breaks the chiral symmetry, gives rise to a splitting of the $N$ degenerate Majorana end states constructed above. With respect to the unnormalized states $\tilde \psi^{(j)}$, this splitting is described by the $N \times N$ matrix
\begin{eqnarray}  \label{eq:eff_Ham}
  \tilde H^{\text{(1)}}_{jl}
  &=& \langle \tilde \psi^{(j)} | H_y | 
  \tilde \psi^{(l)} \rangle
  \nonumber \\ &=&
  \frac{4 i \Delta' m }{W}
  \sum\limits_{nn'} 
  \frac{n n'[1-(-1)^{n+n'}]}{(n'^{2}-n^{2})\sqrt{k_{n}k_{n'}}}\nonumber \\
 && \mbox{} \times
  \sum_{\pm} \left[
  \textrm{Im}\, \frac{ U_{nj} U^{*}_{n'l}}{k_{n'}\pm k_{n}}
  + \frac{2}{ \xi} \textrm{Re}\, 
  \frac{ U_{nj} U^{*}_{n'l}}{(k_{n'} \pm k_{n})^2} 
  \right],
\end{eqnarray} 
where we neglected corrections smaller by a factor of order $(W/\xi)^2$. 
The matrix $\tilde H^{\rm (1)}_{jl}$ is antisymmetric and purely imaginary, which ensures the existence of a single zero-energy bound state if $N$ is odd. In  order to find a true effective Hamiltonian $H^{\rm (1)}$, the eigenvalues of which represent the energies of the fermionic end states, one should transform to the basis of orthogonal states $\psi^{(j)}$ introduced in Eq. \eqref{eq:trafo_orthonorm}, 
\begin{align} \label{eq:effH_orthog}
 H^{\rm (1)} = \frac{1}{\lambda} O^{T} \tilde H^{\rm (1)} O \frac{1}{\lambda}.
\end{align}
In the special case $N=2$, this transformation can be carried out for an arbitrary scattering matrix
$S$ and the energy of the resulting single fermionic bound state is
\begin{align}
  \varepsilon = \frac{|\tilde H^{\rm (1)}_{12}|}{\sqrt{M_{11} M_{22} - M_{12}^2} } .
\end{align}
We now discuss two particular realizations of a metal stub in detail.

\subsection{Rectangular stub}

First, we consider a rectangular stub of length $L$ attached to the spinless $p$-wave superconducting wire, see Fig.\ \ref{fig:3}a. For this geometry, both the scattering matrix $S$ and the Wigner-Smith time-delay matrix $Q$ are diagonal,
\be
  S_{nn'} = -e^{2 i k_{n} L} \delta_{nn'},\ \
  Q_{nn'} = \frac{2 m L}{\hbar k_n} \delta_{nn'},
\ee
with $k_{n}$ given by Eq.\ (\ref{eq:kn}). Since there is no mixing between different channels, the zero energy modes $\tilde \psi^{(j)}$ already form an orthogonal basis. The effective Hamiltonian in the normalized basis $\psi^{(j)} = \tilde \psi^{(j)}/\sqrt{M_{jj}}$ has 
$
  H^{\rm (1)}_{jl} = 0
$
if $ j+l $ is even and
\begin{eqnarray} \label{eq:effH_metal}
  H^{\rm (1)}_{jl} &=&
  \frac{4 i \Delta_{y} \hbar j l}{W(\xi  + 2 L) (l^2-j^2)}
  \sum_{\pm}
  \left\{ \frac{\sin[L  (k_{j} \pm k_{l})]}{k_{l} \pm k_{j}}
  \right. \nonumber \\ && \left. \mbox{}
  + \frac{2 W}{\pi \xi}
  \frac{\cos[L (k_j \pm k_{l})]}{(k_{l} \pm k_{j})^2}
  \right\}
\end{eqnarray}
if $ j+l $ is odd, up to corrections smaller by a factor of order $(W/\xi)^2$.

\begin{figure}[t]
\centering
\includegraphics[width=.4\textwidth]{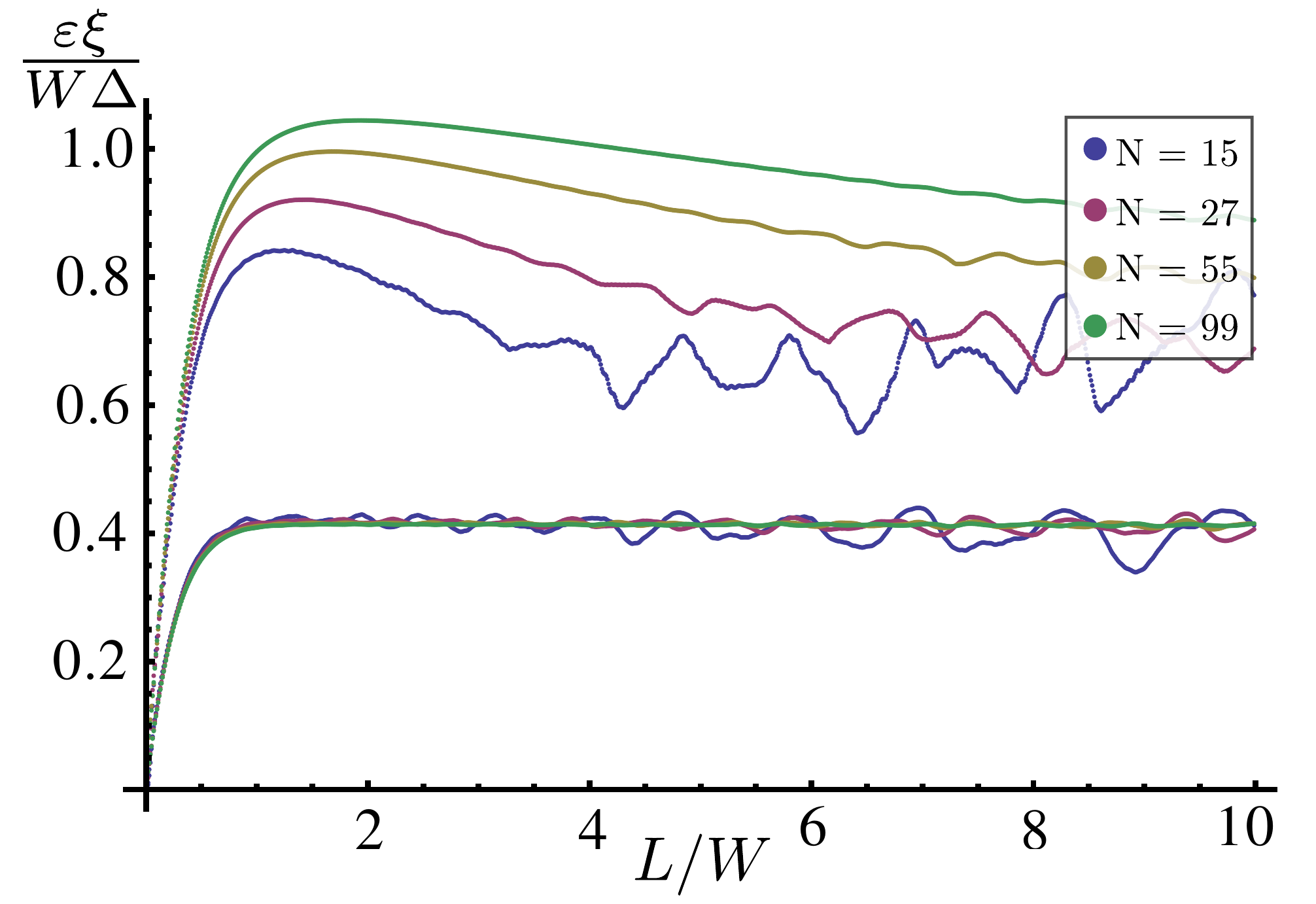}
\caption[]{(Color online) Typical and maximal energies of fermionic subgap states in a spinless $p$-wave superconductor 
with a rectangular normal-metal stub of length $L$ as a function of $L/W$ for different channels numbers 
($ N= 15,\,27,\,55,\,99 $). The maximal energies $\varepsilon_{\rm max}$ have a finite-$N$ correction 
of order $\varepsilon_{\rm m}/\sqrt{N}$, which is why the curves for $\varepsilon_{\rm max}$ still 
show an $N$ dependence for large $N$.}
\label{fig:metal_length}
\end{figure}

The second term in the effective Hamiltonian (\ref{eq:effH_metal}) is smaller than the first one by a factor of order $W/\xi$. However, only this second term contributes in the limit $L=0$ in which there is no normal metal stub.\cite{Kells2012} This is a variation of the cancellation effect pointed out by Potter and Lee.\cite{Potter2012} We now analyze the eigenvalues of the effective Hamiltonian $H^{\rm (1)}$ for finite $L$, when the first term between brackets dominates. 

\begin{figure}
\centering
\includegraphics[width=.5\textwidth]{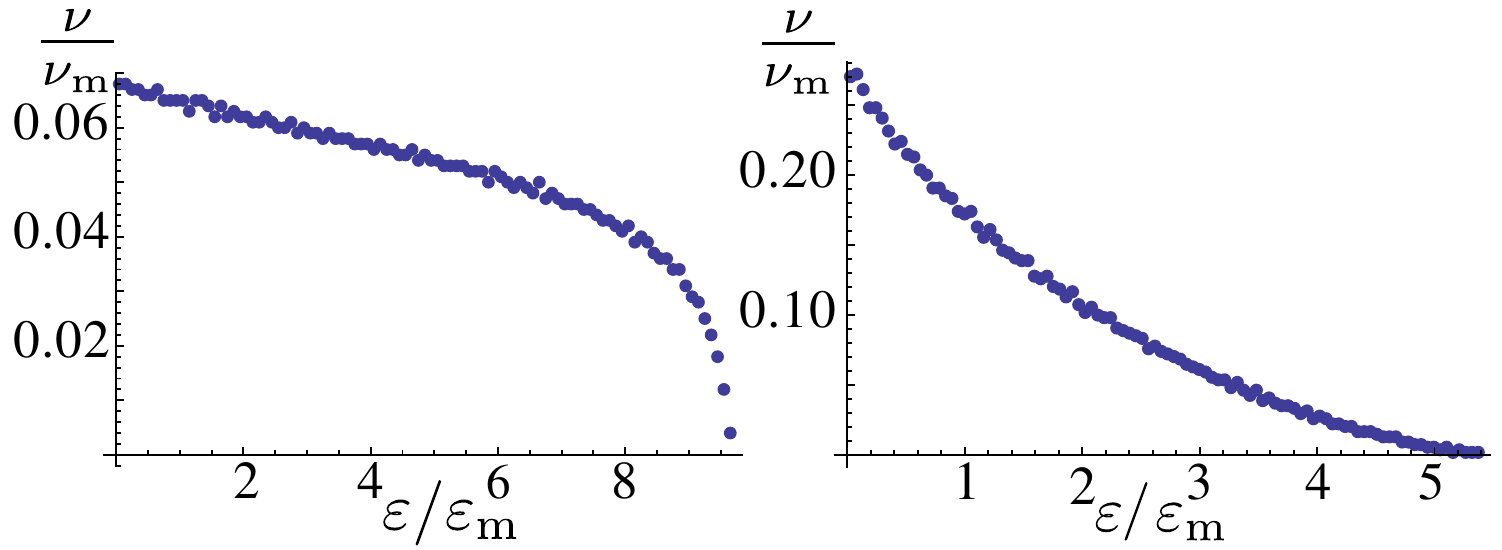}
\caption[]{(Color online) Level density of fermionic subgap states for a rectangular stub in the limit of large $N$, 
for $L/W = 0.1 $ (left) and $L/W = 3 $ (right). The level density is measured in units of $\nu_{\rm m} = 
N/\varepsilon_{\rm m}$.}
\label{fig:levdensmetal}
\end{figure}

Since no closed-form expressions for the eigenvalues of $H^{\rm (1)}$ could be obtained, we numerically diagonalized $H^{\rm (1)}$ and investigated the dependence of the minimum, typical, and maximal positive eigenvalues on the ratio $L/W$ as well as the channel number $N$. 
For $L \sim W$, this analysis gives
\be
  \varepsilon_{\rm typ} \sim \varepsilon_{\rm m} \equiv \frac{W \Delta}{\xi},
  \label{eq:etyp}
\ee
see Fig.\ \ref{fig:metal_length}. The maximum and minimum energies of the subgap states scale as 
$\varepsilon_{\rm min} \sim \varepsilon_{\rm typ}/N$, $\varepsilon_{\rm max} \sim \varepsilon_{\rm typ}$. 
A similar analysis for $k_{\rm F}^{-1} \ll L \ll W$ gives estimates for $\varepsilon_{\rm min}$, 
$\varepsilon_{\rm typ}$, and $\varepsilon_{\rm max}$ which are smaller by a factor $L/W$, 
whereas for $L \ll k_{\rm F}^{-1}$, they are smaller by a factor $L^3 k_{\rm F}^2/W$. A 
crossover to the results of Ref.\ \onlinecite{Kells2012} takes place for $L \lesssim (W^2/k_{\rm F}^2 \xi)^{1/3}$. In the limit of large $N$ the energies of the fermionic subgap states are best described through their level density, which is shown in Fig.\ \ref{fig:levdensmetal}. 

\subsection{Chaotic Cavity}

As a second example, we consider a chaotic cavity attached to the end of the superconducting wire, see Fig.\ \ref{fig:3}b. In this case, the unitary matrix $U$ is randomly distributed in the unitary group,\cite{Beenakker1997} whereas the proper delay times have the probability distribution \cite{Brouwer1997}
\begin{eqnarray}
 P(\tau_{1},...,\tau_{N}) &\propto& \prod_{j=1}^{N} \theta(\tau_{j}) \tau_{j}^{-3N/2-1} e^{-N \bar \tau/2\tau_{j}} 
  \nonumber \\ && \mbox{} \times \prod_{i<j}  |\tau_{i} - \tau_{j} |,
\end{eqnarray}
with the average delay time  $ \bar \tau $. In this case, the matrix $U$ is not diagonal, and the prescription of Eq.\ (\ref{eq:effH_orthog}) needs to be used in order to construct the effective Hamiltonian $H^{\rm (1)}$ for the low-energy subgap states. As in the previous case, we could not obtain closed-form expressions for the energies of the fermionic subgap states and had to resort to a numerical analysis, in which the unitary matrices $U$ were generated according to the Haar measure on the unitary group and the time-delays $\tau_i$ according to the probability distribution given above, following the method described in Ref.\ \onlinecite{Cremers2002}. This analysis gives different results for the limiting cases of a ``small cavity'' and a ``large cavity'', corresponding to the inverse mean dwell time $\hbar/\bar \tau$ large or small in comparison to the superconducting gap $\Delta$.

\begin{figure}
\centering
\includegraphics[width=.5\textwidth]{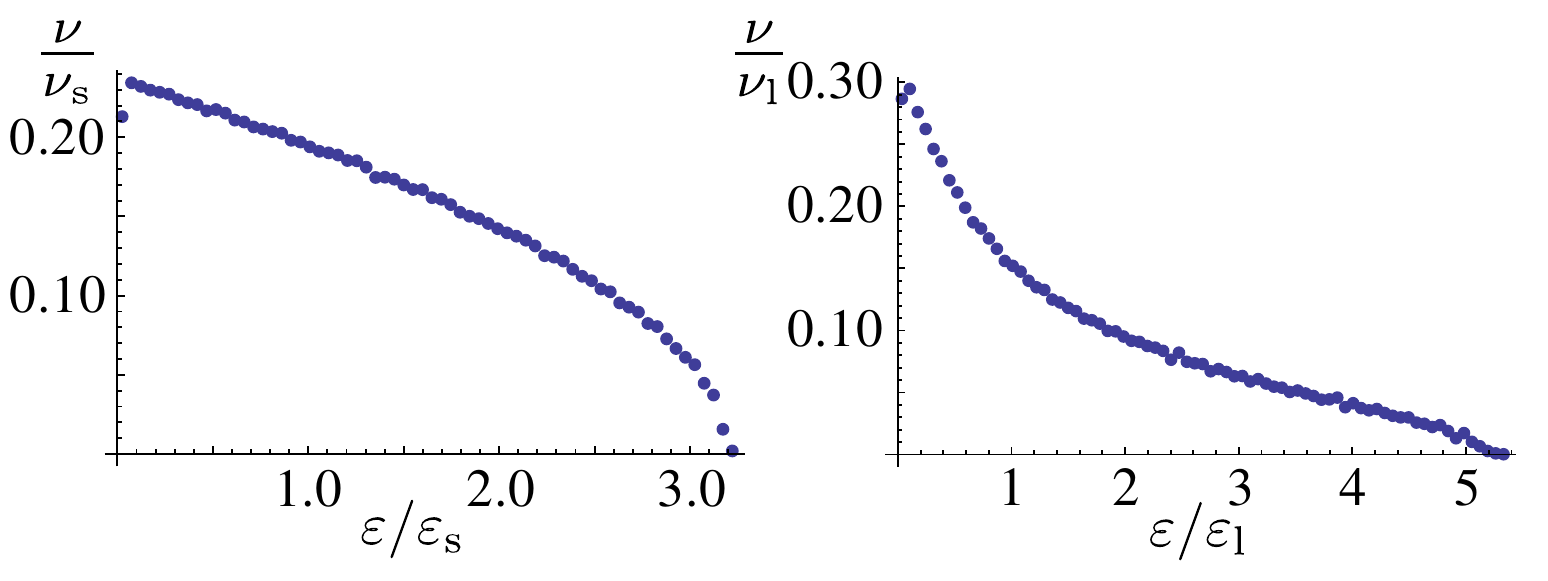}
\caption[]{(Color online) Level density of fermionic subgap states in the small-cavity limit (left) and large-cavity limit (right). The level density is measured in units of $\nu_{\rm s} = N/\varepsilon_{\rm s}$ and $\nu_{\rm l} = N/\varepsilon_{\rm l}$ for the left and right panels, respectively.}
\label{fig:normwire}
\end{figure}

{\em Small-cavity limit.} In the small-cavity limit, the normalization of the $N$ Majorana states $\psi^{(j)}$ is dominated by the integration over the superconducting wire. Not counting the Majorana states, the excitation spectrum of the cavity has a gap comparable to the bulk excitation gap $\Delta$. Upon including $H_y$ one obtains $N$ fermionic subgap states, which have a typical energy 
\be
  \varepsilon_{\rm typ}
  \sim \varepsilon_{\rm s} \equiv \frac{W \Delta}{\xi},
\ee
and $\varepsilon_{\rm max} \sim \varepsilon_{\rm typ}$, $\varepsilon_{\rm min} \sim \varepsilon_{\rm typ}/N$. The precise location of the subgap states depends on the precise scattering matrix of the cavity. The mean level density for an ensemble of cavities is shown in the left panel Fig.\ \ref{fig:normwire}. 

{\em Large-cavity limit.} In the large-cavity limit, the overlap matrix $M_{jl}$ is dominated by the in-cavity parts of the wavefunctions, so that the Majorana modes $\tilde \psi^{(j)}$ are already orthogonal and the effective Hamiltonian $H_{jl}^{\rm (1)} = \tilde H_{jl}^{\rm (1)}/\sqrt{4 \tau_{j} \tau_{l}}$, with $\tilde H_{jl}^{\rm (1)}$ given in Eq.\ (\ref{eq:eff_Ham}). Not counting the Majorana states, the cavity's excitation spectrum has a gap of order $E_{\rm T} = \hbar/\pi \bar \tau$,\cite{Melsen1997} where $\bar \tau$ is the mean dwell time in the cavity. In this case, the typical energy of the fermionic subgap states is
\begin{eqnarray}
\epsilon_{\rm typ} & \sim & \varepsilon_{\rm l} \equiv
  \frac{E_{\rm T} W}{\xi},
\end{eqnarray}
while $\varepsilon_{\rm max} \sim \varepsilon_{\rm typ}$ and $\varepsilon_{\rm min} \sim \varepsilon_{\rm typ}/N$. The mean level density of the subgap states for an ensemble of cavities is shown in the right panel of Fig.\ \ref{fig:normwire}. 

\section{$p+ip$ model with disorder} \label{sec:4}

Whereas strong disorder is known to destroy the topological superconducting phase in the $p+ip$ model
in one dimension, weak disorder with mean free path $l > \xi/2$ preserves the topological 
phase.\cite{Motrunich2001,Brouwer2011b} In this section we investigate the effect of weak disorder on the energies of the fermionic subgap states in a multichannel rectangular $p+ip$ model. Because the disorder is necessarily weak (strong disorder suppresses the topological phase), the effect of disorder can be treated in perturbation theory.

Starting point of our analysis is the chiral-symmetric Hamiltonian $H_0$, which has $N$ normalized Majorana bound states $\ket{\psi^{(j)}}$, $j=1,2,\ldots,N$ at each end of the wire. We take a rectangular geometry, with a wire end and hard-wall boundary conditions at $x=0$, and take the potential $V(\vr)$ to be a Gaussian white noise potential with mean $\langle V(\vr) \rangle = 0$ and variance 
\be
  \langle V(\vr) V(\vr') \rangle = \frac{v_{\rm F}^2}{k_{\rm F} l} \delta(\vr-\vr'),
\ee
where $l$ is the mean free path and $v_{\rm F} = \hbar k_{\rm F}/m$ the Fermi velocity. In our perturbative analysis we treat both the impurity potential $V$ and the transverse superconducting order as perturbations and write
\be
  H = H_0 + U,
\ee
where $U = H_y + H_V$ contains the superconducting correlations coupling to $p_y$ as well as the impurity potential.

The effective Hamiltonian $H_{\rm eff}$ describing the splitting of the $N$ Majorana states into fermionic subgap states can be found using the degenerate perturbation theory of Kato \cite{Kato1949} and Bloch.\cite{Bloch1958} (For additional details on this methodology see also Refs.\ \onlinecite{Messiah1961} and \onlinecite{Jordan2008}.) Defining $P= \sum \ket{\psi^{(j)}}\bra{\psi^{(j)}}$ as the projector onto the zero-energy subspace and $Q=1-P$, we can then write using Bloch's method 
\bea 
  H_{\rm eff} &=& P U P -  P U \frac{Q}{H_0} U P +  P U \frac{Q}{H_0} U \frac{Q}{H_0} U P \non \\ 
&& \mbox{} - \frac{1}{2} \left( P U \frac{Q}{H_0^2} U P U P  + P U P U \frac{Q}{H_0^2} U P \right).
\eea
It is essential to note that the disorder potential $V(\vr)$ alone cannot lift the degeneracy of the Majorana end states at any order of the perturbation theory. This can be understood directly from the observation that the disorder potential $V(\vr)$ does not break the chiral symmetry of the unperturbed Hamiltonian $H_0$ that is responsible for the $N$-fold degeneracy. On the level of perturbation theory this can be understood immediately through the particle-hole symmetry present in the Majorana bound states and the knowledge that for each perturbative diagram that connects Majoranas through the positive energy bulk states there is a cancelling path through the negative energy states. 

Keeping terms to first order in $H_y$ and up to second order in $H_V$ only, we obtain
\bea
H_{\rm eff} &=& H^{(1)}+ H^{(2)}+ H^{(3{\rm a})}-H^{(3{\rm b})},
\eea
with
\bea
H^{(1)}_{jl} &=& \bra{\psi^{(j)}} H_y \ket{\psi^{(l)}},  \non \\
H^{(2)}_{jl} &=& - \bra{\psi^{(j)}} H_y \frac{Q}{H_0} H_V  +  H_V \frac{Q}{H_0} H_y \ket{\psi^{(l)}}, \non \\
H^{(3{\rm a})}_{jl} &=& \bra{\psi^{(j)}} H_y \frac{Q}{H_0} H_V  \frac{Q}{H_0} H_V \ket{\psi^{(l)}} + \text{permutations}, \non \\
H^{(3{\rm b})}_{jl} &=&\frac{1}{2} \sum_k ( V^{(2)}_{jk}  H^{(1)}_{kl} +  H^{(1)}_{jk}  V^{(2)}_{kl} ) ,
\eea
where 
\be
  V^{(2)}_{jl} = \bra{\psi^{(j)}} H_V \frac{Q}{H_0^2} H_V \ket{\psi^{(l)}}.
\ee
The effective Hamiltonian $H_{\rm eff}$ is antisymmetric, which implies that the diagonal elements of all 
the above terms are zero. The first-order term $H^{(1)}$ describes how the transverse superconducting 
correlations lift the degeneracy of the $N$ Majorana modes in the absence of disorder. 
The second-order term $H^{(2)}$ is linear in the disorder potential. 
Its elements are random variables with zero mean and standard deviation that does not appreciably 
change with $\xi$.  The third order term contains two terms, the first of which is also a random 
variable with zero mean and with a root-mean-square proportional $\xi$. 

The term $H^{(3{\rm b})}$ contains corrections to the effective Hamiltonian arising from the renormalization and re-orthogonalization of wavefunctions at the first order of the perturbation theory. Since this term is a weighted sum of first order elements $H^{(1)}_{jl}$, it is the only one of the higher-order perturbation corrections that gives a systematic dependence of energies on the disorder strengths. To see this in more detail, it is instructive to separate the diagonal and the off-diagonal elements of $V^{(2)}$ in the expression for $H^{(3{\rm b})}$,
\bea
  H^{(3{\rm b})}_{jl} &=& \frac{1}{2} (V^{(2)}_{jj}  H^{(1)}_{jl} +  H^{(1)}_{jl}  V^{(2)}_{ll} ) \nonumber \\ 
&& \mbox{} + \frac{1}{2} \sum_{k\ne j}  (V^{(2)}_{jk} H^{(1)}_{kl} + H^{(1)}_{jk}  V^{(2)}_{kl} ). 
\eea
The first term here is the most important because the weights $V^{(2)}_{kk} $ are positive definite random variables. A simple scaling analysis predicts that these variables have both mean and standard deviation proportional to the ratio $\xi/l$ of coherence length and mean free path. This term effectively renormalizes the entire first order contribution, on average driving the energies of the fermionic subgap states towards zero. The second term, which contains the contribution from the off-diagonal elements of $V^{(2)}$, is less important because the disorder potential here connects different Majorana modes. These matrix elements are therefore randomly distributed with zero mean and a root mean square proportional to the coherence length. 

Motivated by these observations, we write the effective Hamiltonian in the form
\be
H \approx \Delta'_y \left[ \left(1- c \frac{\xi}{l} \right)H^{(1)} + H' \right],
\label{eq:Hr}
\ee
where $c = (l/N \xi) \sum_k V^{(2)}_{kk}$ is a number of order unity, and
\be
H' =  H^{(2)}+  (H^{(3 {\rm a})}- \frac{1}{2} \{V^{(2)} - \frac{c \xi}{l}, H^{(1)} \} ).
\label{eq:Ho}
\ee
The correction $H'$ has zero mean. 

\begin{figure}
\centering
\includegraphics[width=.45\textwidth,height=0.35\textwidth]{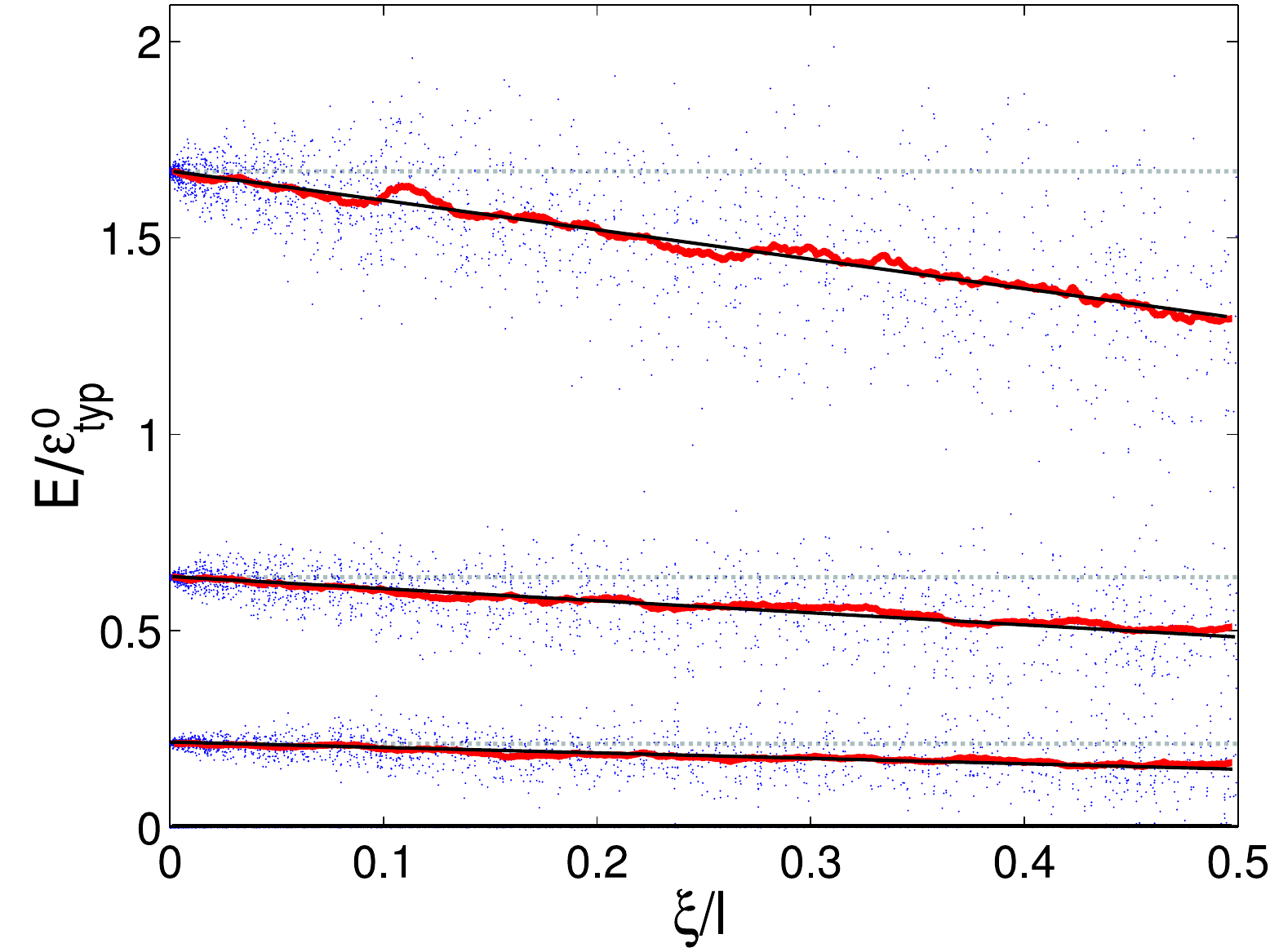}
\caption[]{(Color online) Distribution of energies of fermionic subgap end states in a spinless $p$-wave 
superconductor with $N=7$ channels (dots), as a function of $\xi/l$. For small amounts of disorder the 
term $H^{(2)}$ dominates, pushing the subgap energies up or down with equal probability.
At stronger disorder the term in $H^{(3{\rm b})} \propto \xi/l$ eventually dominates and pulls all 
energy levels towards zero. The red lines indicate the mean calculated from the local distribution of 
eigenvalues. The black lines, which are a linear fit to the mean values in red, share an approximate 
common intercept at the horizontal axis
at $\xi/l = c^{-1} = 2.2$. Dotted grey lines indicate the unperturbed energies.  Energies are measured in units of $\varepsilon_{\text{typ}}^0 = \Delta W^2/\xi^2$. The lattice parameters used in the numerical calculation correspond to $k_F W \approx 23 $ and $k_F 
\xi \approx 320$.} 
\label{fig:Ny28Nx6000}
\end{figure}

We have numerically diagonalized a lattice version of the Hamiltonian (\ref{eq:HBdG}) in order to provide numerical evidence for the applicability of the above arguments. Details of the relationship between the continuum and lattice models can be found in Ref. \onlinecite{Kells2012}. Results of the numerical simulations are shown in Fig.\ \ref{fig:Ny28Nx6000}. For weak disorder, the perturbation $H^{(2)}$ dominates the response of the fermionic subgap states, and the energies of the fermionic subgap states may both increase or decrease, depending on the specific realization of the disorder potential. While large fluctuations persist, for stronger disorder the quadratic-in-disorder perturbation $H^{(3{\rm b})}$ leads to a systematic decrease of the energies of the fermionic subgap states, which is well described by a linear dependence on $\xi/l$, consistent with the first term in Eq.\ (\ref{eq:Hr}).

\section{Smooth potential at wire's end} \label{sec:5}

\begin{figure}
\centering
\includegraphics[width=.45\textwidth,height=0.35\textwidth]{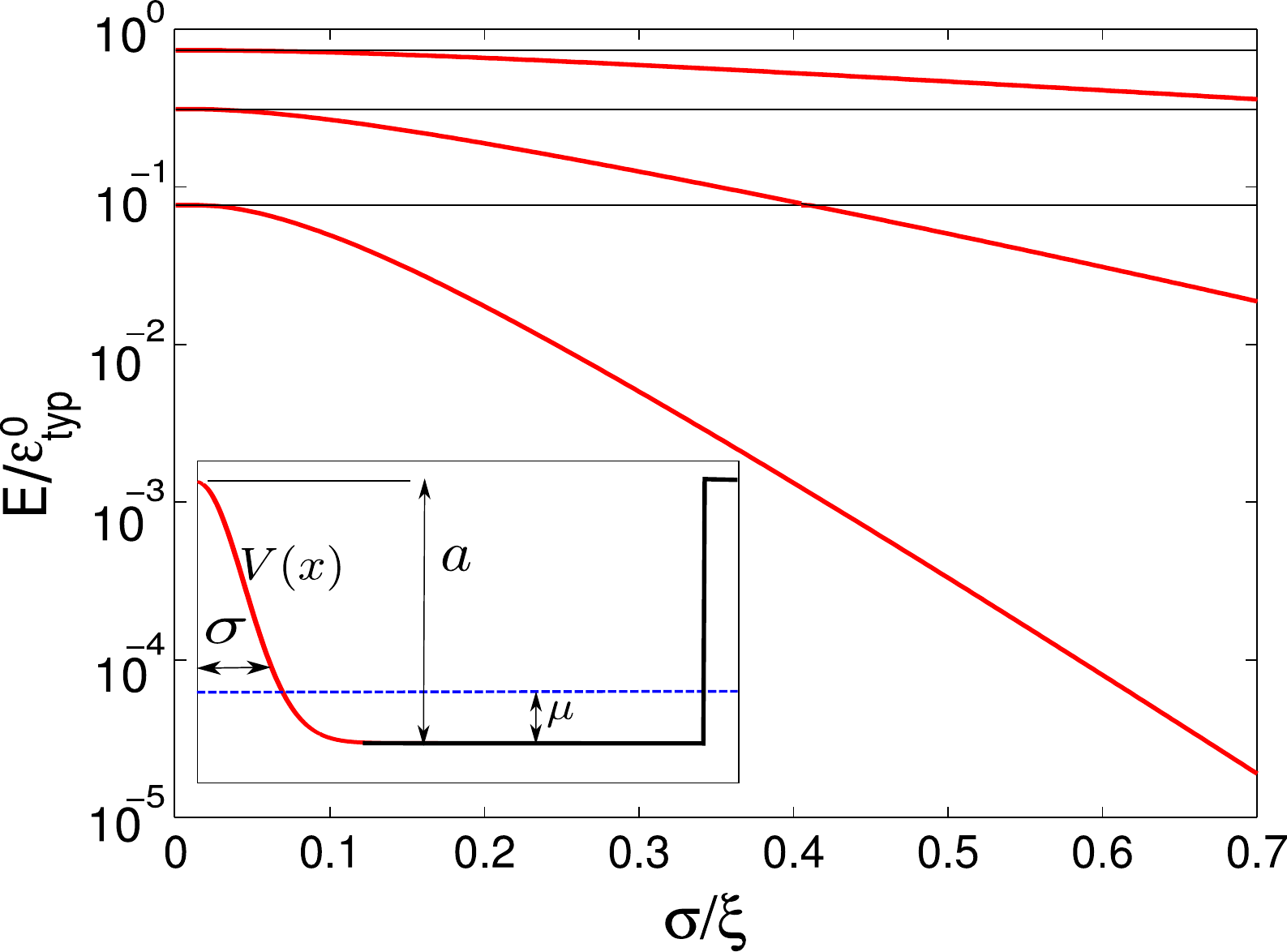}
\caption[]{(Color online) Energies of fermionic subgap end states, in a $6$-channel $p + i p$ wire as function of the adiabaticity parameter $\sigma$ (red). The flat black lines indicate the energies for hard-wall boundary conditions ($\sigma = 0$).  Energies are measured in units of $\varepsilon_{\text{typ}}^0 = \Delta W^2/\xi^2$. The lattice parameters used in the numerical calculation correspond to   $k_F W \sim 19 $ and $k_F \xi \sim 90$, $a = 5 \mu$.  The inset shows the potential profile used in the calculations.}
\label{fig:sigma}
\end{figure}

In this section we consider a wire which is terminated 
by a smooth potential $V(x)$, as shown schematically in the inset of Fig.\ \ref{fig:sigma}. In order to address this scenario we solve the Bogoliubov-de Gennes Hamiltonian in the WKB approximation. Without the transverse pairing term $H_y$ there are $N$ Majorana states $\psi^{(j)}$ with wavefunction
\begin{eqnarray}
  \psi^{(j)}(\vr)  &=& \sqrt{\frac{2 }{W }}
  \left( \begin{array}{c} e^{i\pi/4}\\ e^{-i\pi/4} \end{array} \right) 
  \chi_j(x) \sin \left(\frac{n\pi y}{W}\right), 
\end{eqnarray}
where the functions $\chi_j(x)$ take the form
\begin{eqnarray}
\chi_j(x) &=& \left\{
\begin{array}{ccc} \displaystyle
  \frac{
  e^{-x/\xi + \int_{x_j}^xdx' \kappa_j(x')} }
  {2 \sqrt{\Omega_j \kappa_j(x)}}
  &    \mbox{if $x<x_j$}, \\ \displaystyle
  \frac{e^{-x/\xi} \cos[\Phi_j(x)]}{\sqrt{\Omega_j k_j(x)}}
  &  \mbox{if $x>x_j$},    
\end{array}
\right.
\end{eqnarray}
where $\Phi_j =\pi/4 - \int_{x_j}^x dx' k_j(x') $,
\begin{eqnarray}
\nonumber
k_j(x)&=& \sqrt{2m\left(\mu - V(x) -\frac{\pi^2n^2}{2m W^2}\right)- \frac{1}{\xi^2}}, \\
\nonumber
\kappa_j(x)&=& \sqrt{\frac{1}{\xi^2}-2m\left(\mu -V(x) -\frac{\pi^2n^2}{2m W^2}\right)},
\end{eqnarray}
$\Omega_j$ is the normalization constant, and $x_j$ is the transverse-mode-dependent
classical turning point, defined as the solution of $k_j(x_j)=0 $. 
Inclusion of the transverse pairing Hamiltonian $H_y$ lifts the degeneracy of the $N$ zero energy Majorana end states, where the energy splitting is given by the eigenvalues of the antisymmetric matrix with elements $H^{(1)}_{jl} = 0$ if $j-l$ even and
\begin{eqnarray}
  H^{\rm (1)}_{jl}  &=&  \frac{16 \Delta'}{W}\frac{lj}{j^2-l^2} (X_{jl}^{(1)}+X_{jl}^{(2)}+X_{jl}^{(3)})
\end{eqnarray}
if $j-l$ odd, with, for $j < l$,
\begin{eqnarray}
\nonumber
X_{jl}^{(1)} &=& \int_{-\infty}^{x_j} dx \frac{e^{-2x/\xi-\int_{x_j}^x dx' \kappa_j(x') -\int_{x_l}^x dx' \kappa_l(x')}}{4\sqrt{\Omega_j\Omega_l \kappa_j\kappa_l}}, \\
\nonumber
X_{jl}^{(2)} &=& \int_{x_j}^{x_l} dx \frac{e^{-2x/\xi-\int_{x_l}^x\kappa_l}\cos [\Phi_j(x)]}{2 \sqrt{\Omega_j\Omega_lk_j\kappa_l}} , \\
X_{jl}^{(3)} &=& \int_{x_l}^{\infty} dx \frac{e^{-2x/\xi}\cos[\Phi_j(x)] \cos[\Phi_l(x)]}{\sqrt{\Omega_j\Omega_lk_jk_l}}.
\end{eqnarray}

Figure \ref{fig:sigma} shows numerical calculations for a lattice version of the spinless $p+ip$ model, 
with a potential $V(x)= a e^{-x^2/2 \sigma^2}$, turning the hard-wall ending at $x=0$ effectively into 
a smooth end. The parameter $\sigma$ tunes the length scale over which the potential is turned on. 
The case $\sigma \to 0$ corresponds to a hard-wall boundary. The prefactor $a$ has the dimension of 
energy and determines the height of the potential. For the calculations shown in the figure, we chose 
$a = 5 \mu$. The results of the figure show a clear exponential dependence on $\sigma$ for states on the 
wire end terminated by the smooth function $V$, allowing for energies of the subgap states that are 
significantly below the (already small) estimates for a rectangular geometry with hard-wall boundary 
conditions. 

The origin of the anomalously small energy splittings shown in Fig.\ \ref{fig:sigma} is the smoothness of 
all terms in the Hamiltonian $H$. If the $p+ip$ wire is coupled to a normal-metal stub, as in 
Sec.\ \ref{sec:3}, and the superconducting order parameter $\Delta$ jumps at the interface $x=0$, no 
reduction of the end-state energies is found, even if the normal-metal stub is terminated by a smooth 
potential. (This scenario is well described by the calculation of Sec.\ \ref{sec:3}.) On the other hand, 
one finds a suppression very similar to that shown in Fig.\ \ref{fig:sigma} if the superconducting order 
parameter goes to zero smoothly at the interface. We refere to Ref.~\onlinecite{Kells2012b} for a
discussion of the effect of a smooth confinement in the semiconductor model. 

\section{Conclusion} \label{sec:6}

In this article, we have investigated fermionic subgap states localized near the end of a spinless 
$p$-wave superconducting wires for two terminations of the wire --- a normal-metal stub and a smooth 
confining potential --- and in the presence of weak disorder. The three scenarios give qualitatively 
different estimates for the energies of the subgap states. However, they share the common feature 
that a wire with $N$ transverse channels with a width $W$ that is much smaller than the superconducting 
coherence length $\xi$ has $\mbox{int}\, (N/2)$ fermionic end states, all with energy much below the 
bulk excitation gap $\Delta$. These states appear for the topological phase (which has a Majorana 
fermion at the wire's end), as well as for the non-topological phase (which does not).

The appearance of low-energy fermionic end states poses an obstacle to the identification of 
Majorana fermions through a measurement of the tunneling density of states at the wire's end, 
unless the energy resolution of the experiment is good enough to resolve the splitting between the 
fermionic end states. The corresponding energy scale $\varepsilon_{\rm min}$ scales proportional to 
$\Delta/k_{\rm F} \xi \sim \Delta^2/\mu$ in the most favorable scenario we considered (wire's end 
coupled to a small normal metal stub), which is the same dependence as the subgap states in a vortex 
core.\cite{Caroli1964} The important difference with the subgap states in a vortex core is, however, 
that the number of fermionic end states is limited, so that there exists a maximum energy 
$\varepsilon_{\rm max}$, whereas no such maximum energy exists in a vortex. Other terminations, 
such as a rectangular end with or without disorder, or a smooth confinement potential, give 
significantly smaller values for $\varepsilon_{\rm min}$, and, hence, lead to stricter requirements for the 
energy resolution required to separate an eventual Majorana state from fermionic end states.

The recent experiments that reported the possible observation of a Majorana fermion involved 
semiconductor nanowires with proximity-induced superconductivity.\cite{Mourik2012,Das2012} 
Effectively, the induced superconductivity in these wires is of spinless $p$-wave type. 
However, it should be emphasized that this does not imply that the effective description of such 
a semiconductor wire with $N$ transverse channels is a $p+ip$ model with the same number of 
transverse channels. Instead, only those channels in the semiconductor that are effectively 
spinless ({\em i.e.}, spin polarized or helical, depending on the relative strength of the 
applied magnetic field and the spin-orbit coupling) appear in the effective description in 
terms of a $p+ip$ model. (This latter distinction was overlooked in Ref.\ \onlinecite{Tewari2012}.) 
Typically, this number is smaller than the number 
of transverse channels in the semiconductor. In particular, the nanowires of the experiments 
of Refs.\ \onlinecite{Mourik2012,Das2012} are believed to be thin enough that they map to a 
single-channel $p+ip$ model. Hence, we do not expect that the mechanism for the generation 
of fermionic end states we consider applies to those experiments. However, it will apply to 
nanowires with a larger diameter, which we thus expect to exhibit a clustering of low-energy 
fermionic states in the topologically trivial as well as the topologically nontrivial phases. 
In this context, it is important to note that the condition that $W \ll \xi$ does not a priori prevent 
the applicability of our analysis to thicker wires, because the effective pairing potential
$\Delta$ may decrease with $W$ for proximity-induced superconductivity in the limit of thick wires
(see Ref.\ \onlinecite{Duckheim2011} for an example in which $\Delta \propto W^{-1}$).

We gratefully acknowledge discussions with Felix von Oppen and Inanc Adagedeli. This work is supported by the Alexander von Humboldt Foundation in the framework of the Alexander von Humboldt Professorship, endowed by the Federal Ministry of Education and Research.

\appendix
\section{Relationship between the $p+ip$ and proximity coupled semi-conductor models}

A practical realization of a the $p+ip$ model can be found in semiconducting nanowires with strong spin-orbit coupling, laterally coupled to a standard $s$-wave superconductor and subject to a Zeeman field. In the following we discuss the precise relationship between the models. A related discussion can also be found in  Ref.\ \onlinecite{Alicea2010}.

In two dimensions, and without coupling to the superconductor, the Hamiltonian for this system reads
\begin{eqnarray}
  H_{\rm N} &=& \frac{p^2}{2 m} - \mu + B \sigma_x + \alpha \sigma_y p_x - \alpha' \sigma_x p_y,
\end{eqnarray}
where $\alpha$ and $\alpha'$ set the strength of the spin-orbit coupling and $B > 0$ is the 
Zeeman energy of the applied magnetic field. In the limit of a narrow wire (width $W$ much 
smaller than the coherence length $\xi$ of the induced superconductivity), subgap states as well as
the above-gap quasiparticle states have a vanishing expectation value of the transverse momentum 
$k_y$, which allows us to treat the transverse spin obit term as a perturbation, initially setting 
$\alpha'=0$. Without the term proportional to $\alpha'$ different transverse channels do not couple 
to each other and the eigenfunctions of the Hamiltonian $H_{\rm N}$ are of the form
\be
  \psi_{n,k}^{\pm}(\vr) \propto
  \left( \begin{array}{c}
  e^{-i \theta_k} \\ \pm 1 \end{array} \right)
  \frac{1}{\sqrt{W}} e^{i k x} \sin \frac{\pi n y}{W},
\ee
where the angle $\theta_k$ is defined as
\be
  \sin \theta_{k} = \frac{\alpha k}{\sqrt{B^2 + \alpha^2 k^2}},\ \
  \cos \theta_{k} = \frac{B}{\sqrt{B^2 + \alpha^2 k^2}},
\ee
and the corresponding energies are
\be
  \varepsilon_{k,n}^{\pm} = \frac{\hbar^2 k^2}{2 m} +
  \frac{\hbar^2 \pi^2 n^2}{2 m W^2} - \mu \pm \sqrt{B^2+\alpha^2 k^2}.
  \label{eq:epsilon}
\ee

Upon laterally coupling the semiconductor wire to an $s$-wave superconductor, the excitations are described by the Bogoliubov-de Gennes Hamiltonian
\begin{eqnarray}
  H_{\rm BdG} &=& \left( \begin{array}{cc} H_{\rm N} & \Delta \sigma_y \\
  \Delta \sigma_y & - H_{\rm N}^* \end{array} \right) \nonumber \\ &=&
  \left( \frac{p^2}{2 m } - \mu + B \sigma_x + \alpha \sigma_y p_x \right)
  \tau_z
  \nonumber \\ && \mbox{}
   - \alpha' \sigma_x p_y + \Delta \sigma_y \tau_x,
  \label{eq:1}
\end{eqnarray}
where $\tau_x$, $\tau_y$, and $\tau_z$ are Pauli matrices in electron-hole space. Without the 
transverse spin-orbit coupling $\alpha'$, the Bogoliubov-de Gennes Hamiltonian has a chiral symmetry, 
$\tau_y H \tau_y = -H$. In the basis of the normal-state eigenfunctions $\psi_{n,k}^{\pm}$, 
the Bogoliubov-de Gennes Hamiltonian (\ref{eq:1}) takes the form
\begin{eqnarray}
  H_{\rm BdG} &=& \left( \frac{\hbar^2 k^2}{2 m} +
  \frac{\hbar^2 \pi^2 n^2}{2 m W^2} - \mu \right) \tau_z
  +  \sigma_z \tau_z \sqrt{B^2 + \alpha^2 k^2}
  \nonumber \\ && \mbox{}
  + \Delta \sigma_y \tau_x \cos \theta_k 
  + \Delta  \sigma_z \tau_x \sin \theta_k
   \nonumber \\ && \mbox{}
 - \alpha' p_y (  \sigma_z \cos \theta_k +  \sigma_y \sin \theta_k).
  \label{eq:HDelta}
\end{eqnarray}

In the limit, when both $\Delta$ and the spin orbit energy are smaller than the Zeeman splitting, 
the $s$-wave pairing term proportional to $\sigma_y$ is ineffective, and each transverse channel 
separately maps to two spinless $p$-wave superconductors, one for $\psi_{n,k}^{+}$ and one for $\psi_{n,k}^{-}$.
Neglecting $|\alpha k|$ in comparison to $B$, the corresponding pairing term $\Delta \sin 
\theta_k \sigma_z \tau_x \approx - \Delta' k \sigma_z \tau_x$, with
\be
  \Delta' = -\frac{\alpha \Delta}{B}.
\ee

Without the term proportional to $\alpha'$, the transverse channels in Eq.\ (\ref{eq:HDelta}) can be treated independently (at least in the bulk of the wire, see below). If $\mu < B$, only the ``$-$'' channels (eigenspinors of $\sigma_z$ with eigenvalue $-1$) in Eq.\ (\ref{eq:HDelta}) are topologically nontrivial and can possibly have end states.\cite{Read2000} Projecting the Bogoliubov-de Gennes Hamiltonian in the rotated basis (\ref{eq:HDelta}) onto these channels, one finds an effective Hamiltonian of the form
\begin{eqnarray}
  H_{\rm BdG}^{\rm eff} &=&
  \left( \frac{\hbar k^2}{2 m} + \frac{\hbar^2 \pi^2 n^2}{2 m W^2}
  - \mu - B \right) \tau_z
  \nonumber \\ && \mbox{}
  + \Delta' \hbar k \tau_x
  + \alpha' p_y \, ,
  \label{eq:Hsemispinless}
\end{eqnarray}
Without the transverse spin-orbit coupling $\alpha'$, the effective Hamiltonian (\ref{eq:Hsemispinless}) 
has chiral symmetry and $N$ Majorana end states at each end of the wire. The chiral symmetry is broken 
by the transverse spin-orbit coupling $\alpha'$. 
Because of the particle-hole symmetry of the Majorana modes, the matrix elements of this 
perturbation between the $N$ Majorana end-state of $H_{\rm BdG}^{\rm eff}$ with $\alpha'=0$
are the same as the matrix elements of the $p$-wave superconducting pairing $H_y$ of Eq.\ 
(\ref{eq:HBdG}), if we identify $\Delta' = \alpha'$ in the expression for $H_y$.

If the condition $\mu < B$ is not met, the relation between the semiconductor and $p+ip$ models is 
more complicated. 
For transverse channels for which $\hbar^2 \pi^2 n^2/2 m W^2 < \mu - B$ the wire ends represent a 
chiral-symmetry-preserving perturbation that gaps out eventual Majorana end states, so that such 
channels may be disregarded when considering low-energy end states. For transverse channels for which 
\be
  \mu - B < \frac{\hbar^2 \pi^2 n^2}{2 m W^2} < \mu + B
  \label{eq:cond}
\ee
the Majorana end state in the ``$-$ band'' (eigenspinors of $\sigma_z$ at eigenvalue $-1$ in the 
rotated basis) is protected in the presence of the chiral symmetry, and only perturbations that 
lift the chiral symmetry can lead to a splitting of these end states. Projecting the Bogoliubov-de 
Gennes Hamiltonian in the rotated basis (\ref{eq:HDelta}) onto these channels, one again an effective 
Hamiltonian of the form (\ref{eq:Hsemispinless}),
but with the additional restriction that only those transverse channels that meet the condition 
(\ref{eq:cond}) are considered. The number $N$ of transverse channels that meet this condition may 
be smaller than the original number of propagating channels in the semiconductor.

\end{document}